\documentclass{emulateapj}
\usepackage{psfig}

\newcommand{\ssst}{\scriptscriptstyle}
\newcommand{\E}[1]{\times 10^{#1}}

\newcommand{\RA}[3]{{#1}^{{\rm h}}{#2}^{{\rm m}}{#3}^{{\rm s}}}
\newcommand{\Dec}[3]{{#1}^{\circ}{#2}'{#3}''}

      \newcommand{\ps}{\,{\rm s}^{-1}}
\newcommand{\yr}{\,{\rm yr}}    \newcommand{\Msun}{M_{\odot}}
\newcommand{\cm}{\,{\rm cm}}    \newcommand{\km}{\,{\rm km}}
\newcommand{\parsec}{\,{\rm pc}}\newcommand{\kpc}{\,{\rm kpc}}
        \newcommand{\K}{\,{\rm K}}
    \newcommand{\keV}{\,{\rm keV}}
    \newcommand{\G}{\,{\rm G}}
 
\newcommand{\erg}{\,{\rm erg}}

\newcommand{\no}{n_{\ssst 0}}

\newcommand{\du}{d_{5.2}}

\newcommand{\VLSR}{V_{\rm LSR}}
\newcommand{\twCO}{$^{12}$CO}  \newcommand{\thCO}{$^{13}$CO}
\newcommand{\CeiO}{C$^{18}$O}  \newcommand{\HCOp}{HCO$^+$}
\newcommand{\snr}{{Kes~69}}

\shorttitle{DISCOVERY OF MOLECULAR SHELLS ASSOCIATED WITH SNRs. I. KES 69}
\shortauthors{ZHOU ET AL.}

\begin{document}


\title{DISCOVERY OF MOLECULAR SHELLS ASSOCIATED WITH SUPERNOVA REMNANTS. I. KESTEVEN 69}

\author{Xin Zhou\altaffilmark{1}, Yang Chen\altaffilmark{1}, Yang Su\altaffilmark{1,2}, Ji Yang\altaffilmark{2,3}}

\affil{ $^{1}$ Department of Astronomy, Nanjing University,
Nanjing 210093, China \\
$^{2}$ Purple Mountain Observatory, Chinese Academy of Sciences,
Nanjing 210008, China\\
$^{3}$ National Astronomical Observatories, Chinese Academy of
Sciences, Beijing 100012, China}

\begin{abstract}

Supernova remnant (SNR) \snr\ is morphologically characterized by
brightened radio, infrared, and X-ray emission on the
southeastern rim, with the 1720~MHz OH masers detected in the
northeastern and southeastern regions at various local standard rest (LSR) velocities.
We have performed a millimeter observation in CO and \HCOp\ lines
toward \snr. From the northeastern compact maser region,
\twCO\ and \thCO\ emission's peaks around $65\km\ps$
and $85\km\ps$, which are consistent with the masers' LSR velocities, are detected.
In the southeast, a molecular (\twCO) arc is revealed at
77--$86\km\ps$, well coincident with the partial SNR shell detected
in the radio continuum and mid-infrared observations. An $85\km\ps$ \HCOp\
emission is found to arise from a radio peak on the shell. Both the
molecular arc and the \HCOp\ emission at $\sim85\km\ps$ seem to be
consistent with the presence of extended OH masers along the
southeastern boundary of \snr. The morphology correspondence between
the CO arc and other band emission of the \snr\ shell provides
strong evidence for the association between SNR~\snr\ and the
$\sim85\km\ps$ component of molecular gas.
The multiwavelength emissions along the southeastern shell can be
accounted for by the impact of the SNR shock on a dense, clumpy
patch of molecular gas. This pre-existing gas is likely to be a
part of the cooled debris of the material swept up by the
progenitor's stellar wind.  The association
of SNR~\snr\ with the molecular cloud at the systemic velocity of $\sim85\km\ps$
enables us to place the SNR at a kinematic distance of 5.2~kpc.


\end{abstract}


\section{Introduction}

The progenitors of core-collapse supernovae are most probably formed
in giant molecular clouds (MCs). Due to the short lifetime, they are
not far away from their matrices when they explode. Therefore it is
common that the supernova remnants (SNRs) are located in the
vicinity of MCs and may encounter them in evolution. The
association with MCs often results in irregular morphology of the
SNRs in multiwavelengths, which indicates sophisticated shock
interaction with the inhomogeneous environmental medium. About
20 SNRs have been discovered to be physically interacting with ambient
molecular gas based on the detection of the 1720~MHz OH masers
(Frail et al.\ 1996), which are believed to be a tracer of the shock
interaction with MCs (Lockett et al.\ 1999; Frail \& Mitchell 1998;
Wardle \& Yusef-Zadeh 2002).

SNR~Kesteven~69 is thought to be probably associated with MCs,
because of the OH masers detected toward this remnant; however, the
masers are found at various local standard rest (LSR) velocities and
at various projected locations. Green et al.\ (1997) detected a
compact OH maser using the Very Large Array (VLA) and the Australia
Telescope Compact Array (ATCA) at $\VLSR=69.3\km\ps$, which is
located projectionally in the northeastern part of the remnant.
Recently Hewitt, Yusef-Zadeh, \& Wardle (2008) not only found that
this compact maser also has faint emission at $85\km\ps$, but also
detected extended OH maser emission at the velocity of $\sim85\km\ps$
with the Green Bank Telescope observation and the VLA archival
observation toward the southern bright radio shell. The different
LSR systemic velocities imply the MCs at different distances that
may be impacted by the SNR shock wave. Therefore investigation is
needed to clarify at which systemic velocity the maser emission is
the product of the \snr\ shock interaction.


\snr\ has an irregular X-ray morphology, as observed by {\sl ROSAT}
and {\sl Einstein}, inside an incomplete radio shell (Seward 1990;
Yusef-Zadeh et al.\ 2003). The {\sl Spitzer} Infrared Array Camera (IRAC) mid-infrared
observation toward \snr\ shows an arc at 4.5$\mu$m in the same
location as the southeastern radio shell (Reach et al.\ 2006). The
extended OH emission along the southern incomplete radio and Infrared (IR)
shell hints an interaction of the SNR with the dense molecular gas
in the south. If the SNR/MC association is established, a big
progress can be made toward resolving the open question of the
disparate velocity maser components and the distance to Kes~69 can
also be determined.

Motivated by the supposed association of \snr\ with molecular gas,
we have performed millimeter CO and \HCOp\ observations toward this
remnant. The observations and results are described in \S2 and \S3,
and the conclusion is summarized in \S4.

\section{Observation and Data Reduction}

The observations of millimeter molecular emissions toward SNR~\snr\
were made in two epoches during 2006 November--2007 January and 2007
October--November with the 13.7 m millimeter-wavelength
telescope of the Purple Mountain Observatory at Delingha (hereafter
PMOD). An SIS receiver was used to simultaneously observe the
$^{12}$CO (J=1--0), $^{13}$CO (J=1--0), and C$^{18}$O (J=1--0) lines.
We mapped a $25'\times 25'$ area that contains the full extent of
SNR~\snr\ via raster-scan mapping with the grid spacing of
$30\arcsec$--$60\arcsec$. The main-beam efficiency in the observing
epoch was 67\% and elevation calibration
\begin{center}
\begin{deluxetable*}{cllll}
\tablecaption{Observational parameters\label{tb:obs}}
\tablehead{\colhead{Line}
            &\colhead{frequency (GHz)}
            &\colhead{noise\tablenotemark{a} (K)}
            &\colhead{$\Delta v$\tablenotemark{b} (km $s^{-1}$)}
            &\colhead{FWHM (\arcsec)} }

\startdata
\twCO\ (J=1--0)&115.271204&0.29&0.37 & 60\\
\thCO\ (J=1--0)&110.201353&0.24&0.11 & 60\\
\CeiO\ (J=1--0)&109.782183&0.20&0.12 & 60\\
\HCOp\ (J=1--0)&89.188526&0.04&0.14 & 78\\
\enddata

\tablenotetext{a}{Average rms noise of all final spectra;}
\tablenotetext{b}{Channel separation.}
\end{deluxetable*}
\end{center}
was performed. The
typical system temperature was around 140--280 K. Three
Acousto-Optical spectrometers (AOS) were used as the back end, and the
corresponding spectral coverages were 145 MHz for $^{12}$CO (J=1--0)
and 43 MHz for both $^{13}$CO (J=1--0) and C$^{18}$O (J=1--0), all
divided into 1024 channels. The observed velocity ranges were
$-120$--$(+260)\km\ps$ for $^{12}$CO, 11--$127\km\ps$ for $^{13}$CO
(J=1--0), and 10--$128\km\ps$ for \CeiO\ (J=1--0). We also chose two
points for the \HCOp\ (J=1--0) line observation with long-time
integration [122 minutes for point ($\RA{18}{32}{50}$,
$\Dec{-10}{12}{42}$) and 134 minutes for ($\RA{18}{33}{10}$,
$\Dec{-10}{12}{42}$)], with the velocity range $-3$ to $+141\km\ps$.
The baseline subtraction was performed with low-order polynomial fit.
In Table~\ref{tb:obs} we list some observational parameters, such as
the frequencies, the average rms noises of all final spectra, the
channel separations, and the FWHM. All
data were reduced using the GILDAS/CLASS
package\footnote{http://www.iram.fr/IRAMFR/GILDAS}.

The H{\footnotesize I} line emission data of the archival VLA Galactic Plane Survey
(VGPS; Stil et al.\ 2006) were also processed. The observations
were made by the VLA and the Green Bank
Telescope, presented with the beam-size of $1'$, the velocity
resolution of $1.56\km\ps$, and the rms noise of 2~K per
$0.824\km\ps$ channel. The processed {\sl Spitzer} IRAC (Fazio et al.\ 2004) Basic Calibrated Data 
were used here (PID: 146, PI: Ed Churchwell), which are available
in the archive of the {\sl Spitzer} Science Center
and include flat-field correction, linearity
and flux calibrations, and dark subtraction. The final mosaic $4.5$ 
$\mu$m IR image was produced by further processing with the custom
IDL software (Huang et al. 2004).
We also used the archived {\sl Spitzer} Multiband Imaging Photometer
(MIPS; Rieke et al.\ 2004) Post Basic Calibrated Data to present the
mid-IR $24$ $\mu$m image, which were obtained from the Micron Survey
of the Inner Galactic Disk Program (PID: 20597, PI: Sean Carey).
The 1.4~GHz radio continuum emission data were obtained from
the NRAO VLA Sky Survey (NVSS; Condon et al.\ 1998).

\section{Results}

\subsection{The CO and \HCOp\ Emissions}

\begin{figure}[tbh!]
\centerline{ {\hfil\hfil
\psfig{figure=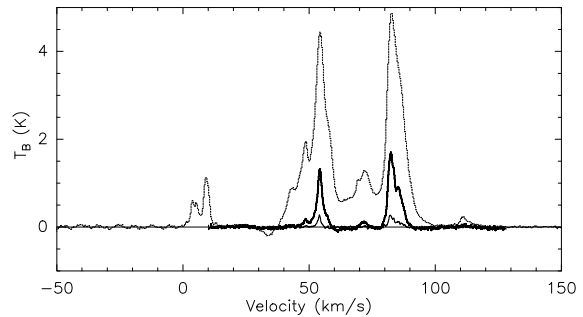,height=2.3in,angle=-90, clip=} \hfil\hfil}}
\caption{The spectra of the central point ($\RA{18}{32}{50}.0,
\Dec{-10}{06}{42}$) of the field of view in our observation
toward Kes~69: dashed line for $^{12}$CO, thick solid line
for $^{13}$CO, and thin solid line for C$^{18}$O. The S/N 
is highest here. The small dip near $33\km\ps$ results
from the weak emission at the reference position, which is
far from the concerned velocities.} \label{f:00spec}
\end{figure}

Figure~\ref{f:00spec} shows three CO spectra toward the center
($\RA{18}{32}{50}.0,$ $\Dec{-10}{06}{42}$) of the mapping area.
There are several velocity components in the velocity range 0--$120
\km\ps$ in the $^{12}$CO (J=1--0) spectrum, and no \twCO\ emission is
detected out of this range in the whole mapping area. The \twCO\
emission peaks appear in the intervals $\sim2$--12, 40--60, 60--76,
76--100, and 107--$120\km\ps$. We have found no morphological
correspondence between the CO emission in the velocity intervals
2--$12\km\ps$ and 40--$60 \km\ps$ and the emission from \snr\ in any
other waveband. The molecular component at around $110 \km\ps$ also
shows no morphological correspondence, which is near the tangent
point in this direction. Applying the rotation curve of Clemens
(1985) together with $R_{0}=8.0 \kpc$ (Reid 1993) and $V_{0}=220
\km\ps$, the tangent point in this direction is at 7.4 kpc at $113.7
\km\ps$. For the \twCO\ emission, we focused our analysis on the
velocity intervals 60--76$\km\ps$ and 76--100$\km\ps$, which respectively
cover $69.3\km\ps$ and $85\km\ps$ at which the OH masers were
detected. The \thCO\ (J=1--0) emission is prominent at around
$54\km\ps$ and 82$\km\ps$, at which the \twCO\ emission is strong. The
intensity of the C$^{18}$O (J=1--0) emission is too weak to examine its
spatial distribution.

\begin{figure*}[tbh!]
\centerline{ {\hfil\hfil \psfig{figure=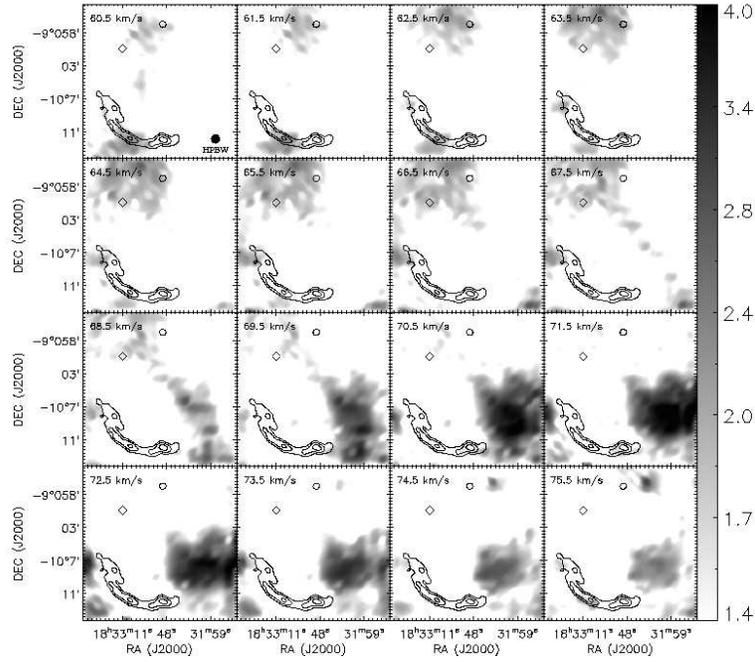,height=4.2in,angle=0,
clip=} \hfil\hfil}} \caption{$^{12}$CO J=1--0 emission maps
integrated each 1 km s$^{-1}$ (smoothed to an angular resolution of
$0'.4$ by linear interpolation), with the NVSS 1.4 GHz radio
continuum emission in contours (contour levels are 20, 49, and 107
mJy beam$^{-1}$). Central velocities are indicated in each image. The rms
noise of each map is $\sim0.18$ K~km~s$^{-1}$ and the beam size is
1\arcmin. The diamond denotes the location of the compact maser at
$69.5\km\ps$.} \label{f:12co6076}
\end{figure*}

We have produced $^{12}$CO intensity maps with interval $1\km\ps$
between 60 and $76\km\ps$ (Fig.~\ref{f:12co6076}) to examine the
molecular gas around 69.3 $\km\ps$, at which the compact OH maser is
seen.  In the northeastern compact maser region,
some faint diffuse $^{12}$CO emission are present at 60--$70\km\ps$.
A distinct cloud at 69--76$\km\ps$ is seen in the west.

\begin{figure*}[tbh!]
\centerline{ {\hfil\hfil \psfig{figure=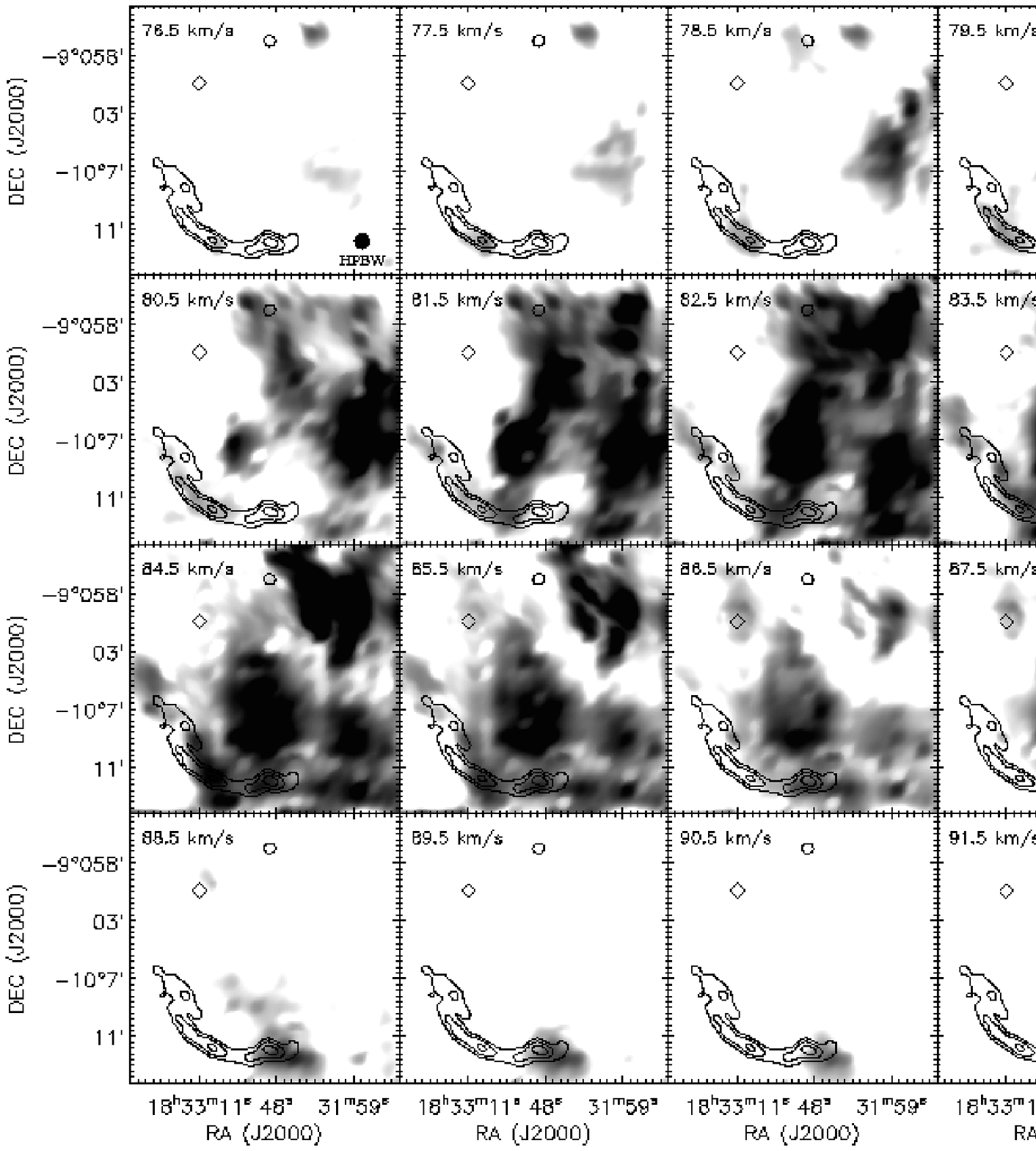,height=4.2in,angle=0,
clip=} \hfil\hfil}}
\caption{The same as for Figure~\ref{f:12co6076},
but different velocity. } \label{f:12co7692}
\end{figure*}

\begin{figure}[tbh!]
\centerline{ {\hfil\hfil \psfig{figure=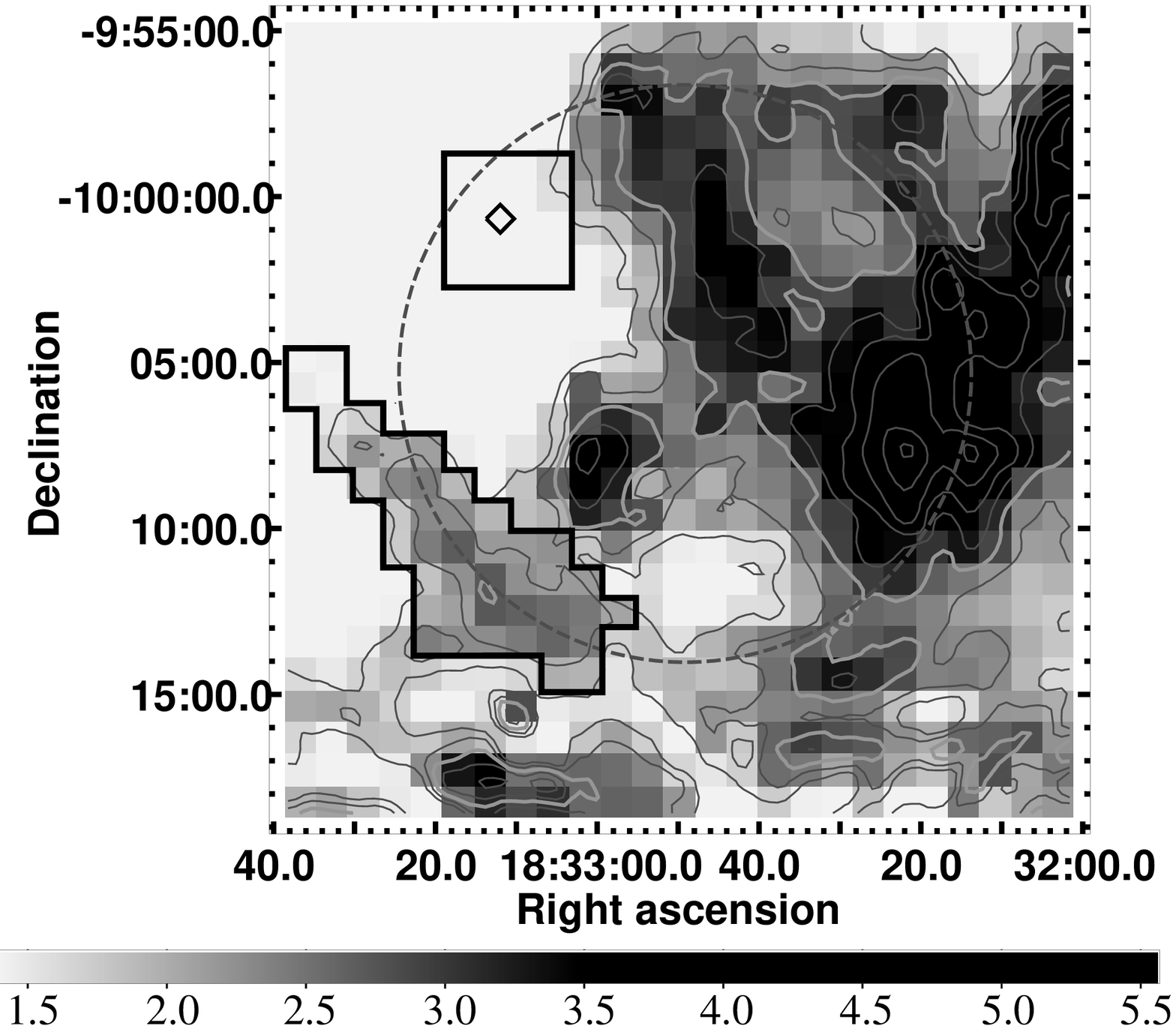,height=2.8in,angle=0,
clip=} \hfil\hfil}} \caption{\twCO\ (J=1--0) intensity map in the
velocity interval 80--$81\km\ps$. The intensity contours (smoothed
to a resolution of $0'.24$ by interpolation) are at linear scale
levels from 1.67 to 5.57 K$\km\ps$ in steps of 0.6 K$\km\ps$, and
the thick contour denotes the half-maximum intensity (2.79 K$\km\ps$). 
The rms noise is $\sim0.18$ K$\km\ps$. The dashed circle
is plotted roughly running through the molecular arcs in the
southeast and the northwest, with an angular radius of $8'.7$. The
diamond denotes the location of the compact OH maser at
$69.3\km\ps$. Two regions for CO spectrum extraction that overlap
the northeastern OH maser and southeastern CO arc are also shown.
}\label{f:COregion}
\end{figure}

\begin{figure}[tbh!]
\centerline{ {\hfil\hfil \psfig{figure=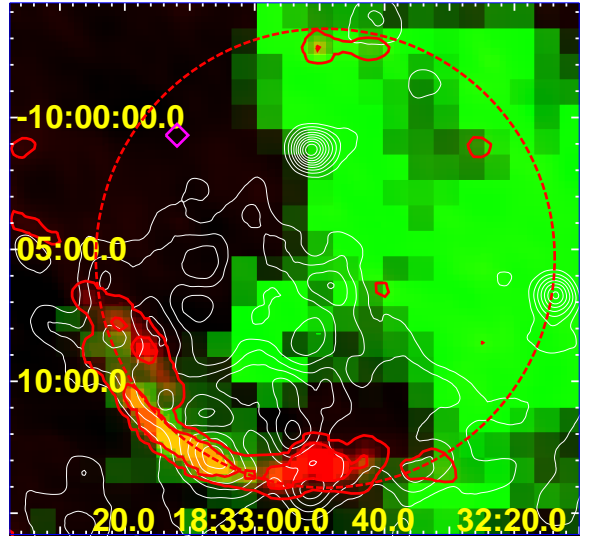,height=2.8in,angle=0,
clip=} \hfil\hfil}}
\caption{The \twCO\ (J=1--0) intensity map in the velocity interval
79--$82\km\ps$ (above $5\sigma$) coded in green,
overlaid with the {\sl ROSAT} PSPC X-ray contours in white, and the
NVSS radio continuum emission in red (the contour levels are 10, 41,
and 172 mJy beam$^{-1}$). The diamond denotes the location of the compact
OH maser at $69.3\km\ps$. The dashed circle is plotted roughly
running through the molecular arcs in the southeast and the
northwest, with an angular radius $8'.7$. } \label{f:3c}
\end{figure}

We have also produced $^{12}$CO intensity maps over the velocity
range of 76--92 $\km\ps$ with interval $1 \km\ps$
(Fig.~\ref{f:12co7692}).By comparison with the radio continuum
emission, an arc of the molecular gas at 77--$86\km\ps$ is
strikingly coincident along the southeastern rim. This arc is
clearly seen in the close-up intensity images at 80--81$\km\ps$
(Fig.~\ref{f:COregion}) and 79--$82\km\ps$ (Fig.~\ref{f:3c}).
In the northwest (see Figs.~\ref{f:COregion}
and~\ref{f:3c}), there is another section of molecular arc, which
can even be discerned in the wider range 79--$86\km\ps$
(Fig.\ref{f:12co7692}). The two sections of molecular arcs are seen
in similar velocity range and can be threaded with a circle of
angular radius $8'.7$. Some bright, complicated \twCO\ features are
present in the field of view at 80--$88\km\ps$. There may be a
contribution from the H{\footnotesize II} region G21.902--0.368 in the northwest,
where the recombination line at 79.5$\km\ps$ has been detected (Lockman 1989).
There is also a molecular cloudlet at 87--$91\km\ps$,
coincident with the strongest southern radio peak, at the western
end of the incomplete radio shell. If this small cloud is associated
with the SNR, the radio peak can be accounted for by the impact of
the remnant shock on it, because of the drastic shock deceleration
and hence the magnetic field compression and amplification.


\begin{center}
\begin{deluxetable*}{cllll}
\tablecaption{Fitted and derived parameters for the MCs around $85\km\ps$
in the southeastern region\tablenotemark{a}\label{tb:SE}}

\tablehead{ \multicolumn{5}{c}{\sl Gaussian Components}\\
\tableline
             \colhead{Line}
            & \colhead{Center (km~s$^{-1}$)}
            & \colhead{FWHM (km~s$^{-1}$)}
            & \colhead{$T_{\rm peak}$ (K)}
            & \colhead{$W$ (K km s$^{-1}$)}}

\startdata
\twCO\ (J=1--0) &84.7&5.2&2.4&13.4\\
\thCO\ (J=1--0) &84.9&2.4&0.5&1.2\\
\tableline\tableline
\multicolumn{5}{c}{\sl Molecular Gas Parameters}\\
\tableline
& $N({\rm H}_2)$ $(10^{21}\cm^{-2})$\tablenotemark{b}
& $M(\Msun)$\tablenotemark{b}
& $T_{\rm ex}$~(K)\tablenotemark{c}
& $\tau(^{13}$CO)\tablenotemark{d}\\
\tableline
Gaussian components: & 2.4 / 1.6 & $5.8\E{3}\du^2$ /
$3.8\E{3}\du^2$\,\tablenotemark{e} & 7.8 & 0.2\\
Residual part\tablenotemark{f} : & 2.2 & $5.2\E{3}\du^2$\,\tablenotemark{e} & & \\
The whole\tablenotemark{g} : & 4.6  & $1.1\E{4}\du^2$\,\tablenotemark{e} & & \\

\enddata

\tablenotetext{a}{The region is defined in Figure~\ref{f:COregion}.}

\tablenotetext{b}{See text for the two estimating methods.}


\tablenotetext{c}{The excitation temperature calculated from the
maximum \twCO (J=1--0) emission point in the region.}

\tablenotetext{d}{The optical depth of the \thCO\ (J=1--0) line.}

\tablenotetext{e}{$\du=d/(5.2\kpc)$ (see \S\ref{dynamics}).}

\tablenotetext{f}{Determined by subtracting the Gaussian components
centered at $72\km\ps$ and $85\km\ps$ after the multiple Gaussian
fitting for the \twCO\ (J=1--0) emission in the velocity range
70--$95\km\ps$.}

\tablenotetext{g}{Combination of the Gaussian components around
$85\km\ps$ and the residual part.}
\end{deluxetable*}
\end{center}

\begin{center}
\begin{deluxetable*}{cllll}
\tablecaption{Fitted and derived parameters for the \HCOp\ emitting point
($\RA{18}{33}{10}.0$, $\Dec{-10}{12}{42}$) \label{tb:HCO}}

\tablehead{\multicolumn{5}{c}{\sl Gaussian Components}\\
\tableline
            \colhead{Line}
            & \colhead{Center (km~s$^{-1}$)}
            & \colhead{FWHM (km~s$^{-1}$)}
            & \colhead{$T_{\rm peak}$ (K)}
            & \colhead{$W$ (K km s$^{-1}$)}}

\startdata
\twCO\ (J=1--0) &85.0&3.4&3.9&14.3\\
\thCO\ (J=1--0) &84.9&2.4&0.5&1.2\\
\HCOp\ (J=1--0) &85.0&2.2&0.2&0.5\\
 \tableline\tableline
\multicolumn{5}{c}{\sl Molecular Gas Parameters}\\
\tableline
$N(^{13}{\rm CO})$ $(10^{15}\cm^{-2})$ & $N({\rm HCO}^{+})$
$(10^{11}\cm^{-2})$
 & $\tau(^{13}$CO)& \\
\tableline
1.3 & 3.5 & 0.13&\\
\enddata
\end{deluxetable*}
\end{center}

\begin{figure}[tbh!]
\centerline{ {\hfil\hfil
\psfig{figure=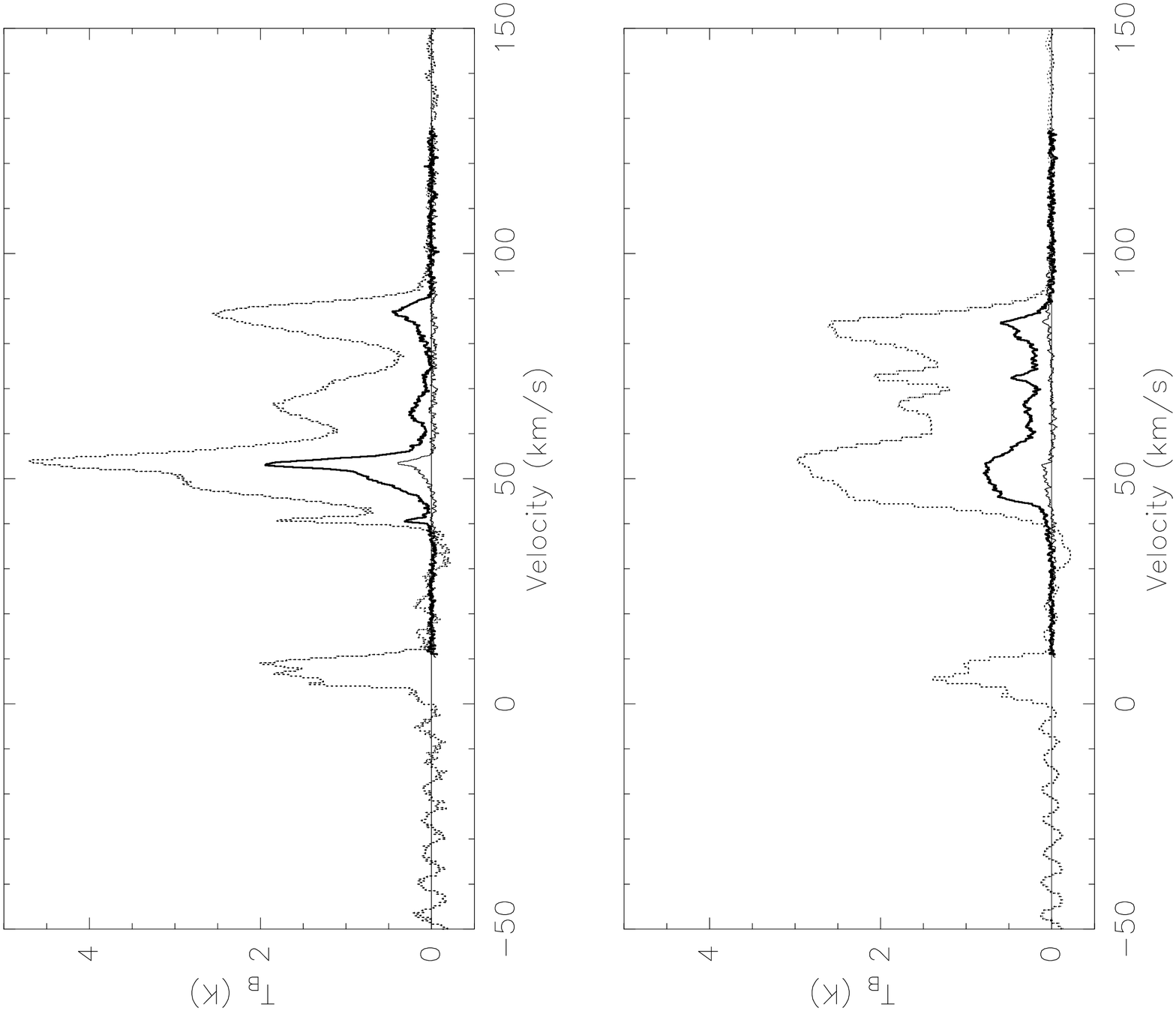,height=4.3in,angle=-90,clip=}}} \caption{The CO
spectra from the two regions that are defined in
Figure~\ref{f:COregion}. The upper panel is for the northeastern
region covering the compact maser and the lower panel for the
southeastern shell region. The dashed lines stand for the $^{12}$CO
emission, the thick solid lines for $^{13}$CO, and the thin solid
lines for C$^{18}$O. The sinusoidal pattern above the noise limit
that appears in the \twCO\ spectra for negative LSR velocities
originates from the nonlinear response of the AOS system. }
\label{f:regionspec}
\end{figure}
The CO spectra from the northeastern compact OH maser region and the
southeastern shell region are shown in Figure~\ref{f:regionspec}
(the two regions for CO spectrum extraction are shown in
Figure~\ref{f:COregion}). For the former region, the \twCO\ and
\thCO\ emissions peak at both $\sim65$ and $85\km\ps$
(both with a signal-to-noise ratio (S/N)$>$3), essentially in
agreement with the velocities at which the single maser arises
(Green et al.\ 1997; Hewitt et al.\ 2008). The \twCO\ and \thCO\
lines at $\sim65\km\ps$ have complicated profiles, each of which can
be phenomenologically fitted with a combination of at least three
Gaussians. The lines at $85\km\ps$ are slightly broadened in the
blue wings. For the latter region, there are \twCO\ and \thCO\ lines
at $85\km\ps$, with blue wings broadened (although the wing may be
contaminated by the $72\km\ps$ peaks). The profile of each line is
divided into a Gaussian at $85\km\ps$ and a residual part in the
blue wing, with the fitted and derived parameters summarized in
Table~\ref{tb:SE}.

In the derivation, we used two methods to estimate the H$_{2}$
column density and molecular mass. In the first method, the H$_2$
column density is estimated by the use of the conversion factor $N({\rm
H}_2)/W(^{12}$CO)$\approx1.8\times 10^{20} \cm^{-2} \K^{-1} \km^{-1}
{\rm s}$ (Dame et al.\ 2001). In the second method, we assume local
thermodynamical equilibrium (LTE) for the gas and optically thick
condition for the $^{12}$CO (J=1--0) line and use the relation
$N({\rm H}_2)\approx7\E{5}N(^{13}$CO) (Frerking et al.\ 1982).

There seems to be a weak feature at $82\km\ps$ in the \thCO\ line
profile of the southeastern region,
which could be either a broadened part of the $85\km\ps$
component or a chance coincidence of an irrelevant, unperturbed
cloud in the line of sight. We performed a multiple Gaussian fit
incorporating this feature and derived a molecular gas mass for this
feature, which is about two orders lower than the virial mass,
indicating that the cloud is perturbed. In view of this and the
clear shell-like appearance at 80--$82\km\ps$, we treat it simply as
a small part in the broadened wing of the $85\km\ps$ gas.

\begin{figure}[tbh!]
\centerline{ {\hfil\hfil
\psfig{figure=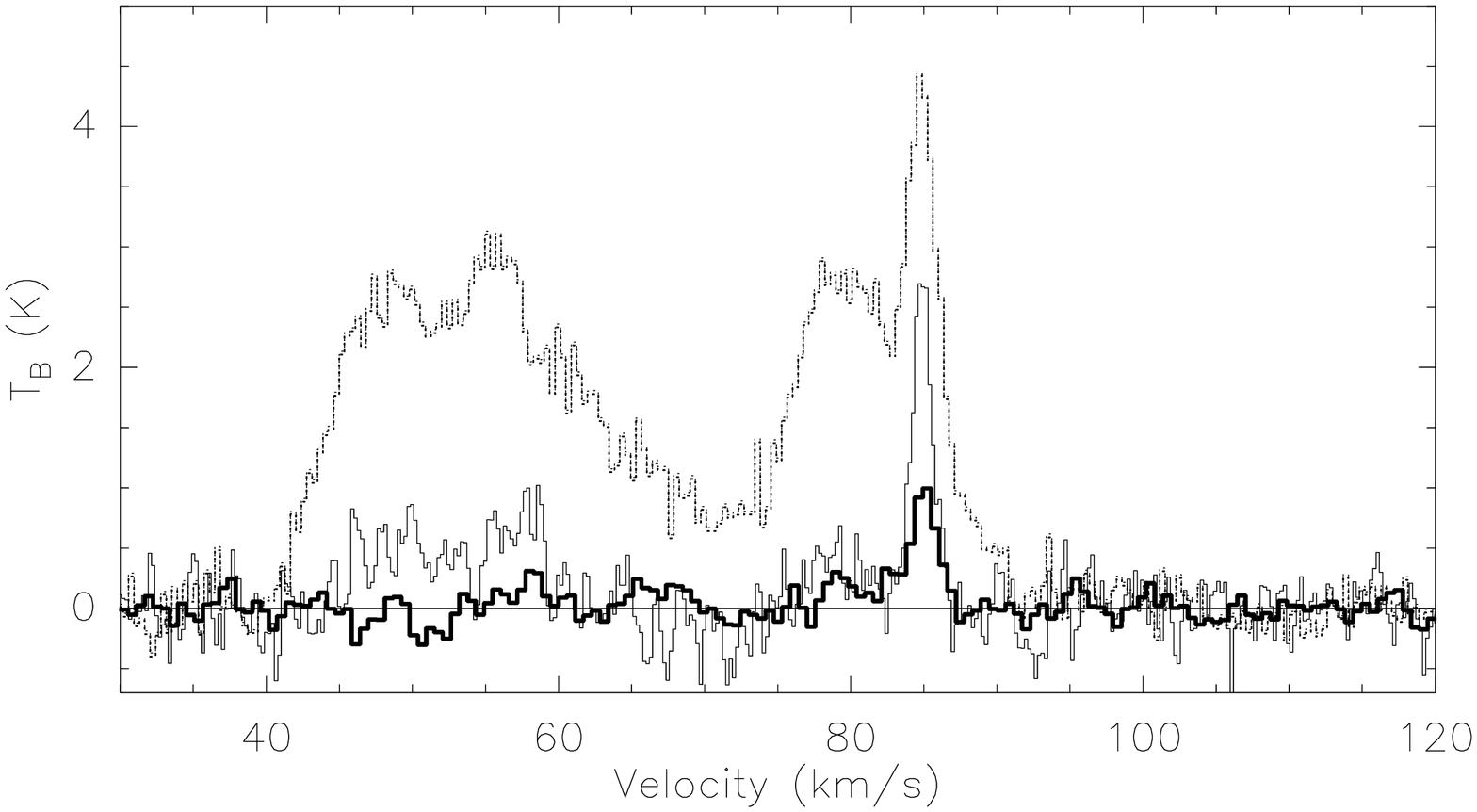,height=2.5in,angle=-90, clip=} \hfil\hfil}}
\caption{The spectra of the point ($\RA{18}{33}{10}.0$,
$\Dec{-10}{12}{42}$) on the southeastern rim of \snr: the dashed line
for $^{12}$CO, the thin solid line for two times the intensity of
$^{13}$CO, and the thick solid line for five times the intensity of \HCOp.
The noise level for the \HCOp\ emission is given in
Table~\ref{tb:obs}, together with those for the CO lines.}
\label{f:HCO}
\end{figure}
In the long-time pointing observation toward the southern strongest
radio peak or the cloudlet mentioned above, which targets at
($\RA{18}{32}{50}.0$, $\Dec{-10}{12}{42}$), no \HCOp\ signal was
detected. However, \HCOp\ emission was detected in the pointing
observation toward another radio peak at ($\RA{18}{33}{10}.0$,
$\Dec{-10}{12}{42}$) on the southeastern shell. The \twCO, \thCO,
and \HCOp\ spectra of this point are shown in Figure~\ref{f:HCO}.
The \HCOp\ emission appears to be prominent only at $\sim85\km\ps$,
at which both the $^{12}$CO and $^{13}$CO emission peak, too. We
calculated the column density of \HCOp\, assuming that the \HCOp\
emission is in LTE with the same excitation temperature as the
\twCO\ emission ($\sim 7.8$~K) and is optically thin.
For comparison with the
CO column density, we smoothed the CO (J=1--0) data to the same
angular resolution as the \HCOp\ observation,
and the results are listed in
Table~\ref{tb:HCO}. Considering the $^{12}$CO$/^{13}$CO ratio
(30--70) of the general interstellar medium (Langer 1992 and
references therein), the \HCOp/CO abundance ratio is $\sim$
5$\E{-6}$. It is somewhat less than that of undisturbed cold MCs
such as TMC-1 or L134N ($1\E{-4}$; Ohishi, Irvine, \& Kaifu 1992).
The \HCOp\ abundance is expected to be reduced in a slow,
nondissociating shock, unless some enhanced ionization is present
(Iglesias \& Silk 1978; Elitzur 1983).  We note that the extended
maser emission on the southern rim is found at the same LSR velocity
(Hewitt et al.\ 2008), which implies that the dense molecular gas at
this velocity is disturbed by C-type shock. For the origin of \HCOp\
emission, though, further observation is needed to make a decision.

\subsection{Morphological correlation in multiwavelengths}
\label{sec:morph}

Combination of the multiwavelength observations toward \snr\
(Fig.~\ref{f:3c}) shows some interesting morphological correlation.

\begin{figure}[tbh!]
\centerline{ {\hfil\hfil
\psfig{figure=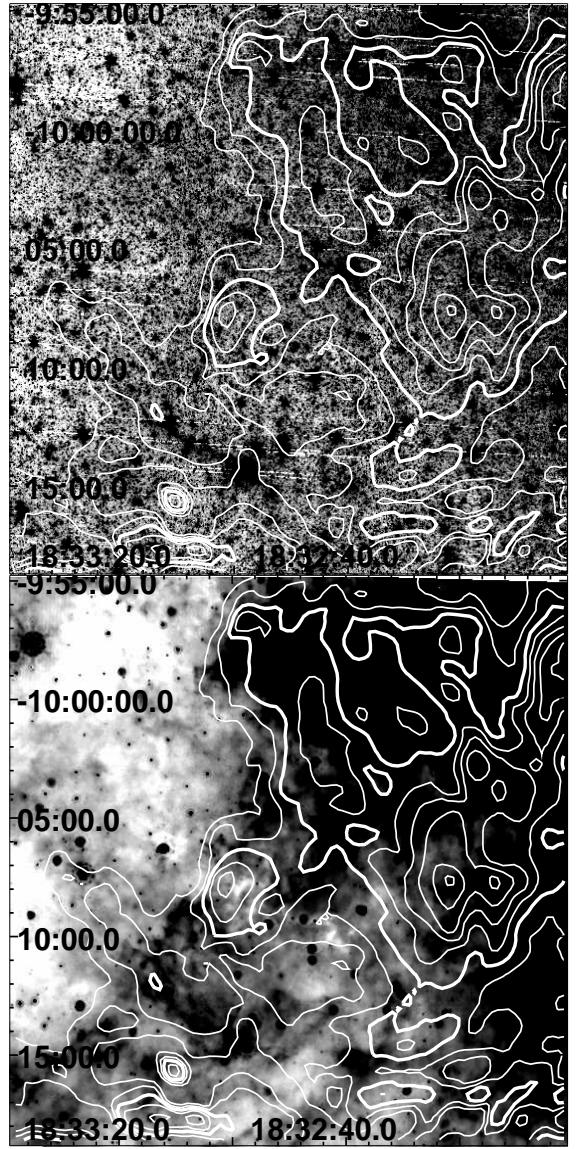,height=6.0in,angle=0,clip=}}} \caption{{\sl
Spitzer} $4.5$ $\mu$m (upper panel) and $24$ $\mu$m (bottom panel)
grey scale images overlaid with the 80--$81\km\ps$ \twCO\ (J=1--0)
intensity contours (the same as in Fig.~\ref{f:COregion}).}
\label{f:IR}
\end{figure}
First, the southeastern 77--$86\km\ps$ molecular shell is not only
coincident well with the 1.4~GHz radio shell (as pointed out above),
but also (as seen in Fig.~\ref{f:IR}) with the mid-IR ``ridge''
described in Reach et al.\ (2006). This mid-IR ridge is prominent in
4.5$\mu$m, with a relatively high brightness in 5.8$\mu$m. A part of
the $24$ $\mu$m floclike emission in the southeast is also aligned
with the molecular shell (Fig.~\ref{f:IR}). The presence of the
molecular shell is consistent with the \HCOp\ and extended 1720~MHz
OH maser emission at $85\km\ps$ along the southern rim. This
correspondence in morphological features is most probably another
signature of the SNR-MC interaction, in addition to the 1720 MHz OH
masers detected in \snr.

Second, as described above, both the sections of molecular arcs at
$\sim77$--$86\km\ps$ in the southeast and the northwest seem to be
aligned along a circle. A faint 1.4~GHz bar in the north
(highlighted by a red contour in Fig.~\ref{f:3c}) is also along this
circle. Moreover, as noted by Yusef-Zadeh et al.\ (2003), there is a
faint radio shell in the northwest in the low-resolution 330~MHz
image of \snr\ (Kassim 1992), which is confused with the H{\footnotesize II} region
G21.902--0.368. We note that the peak of this patch of radio
emission roughly coincides with the northwestern section of the molecular
arc.

These morphological correlations demonstrate that
\snr\ is associated with the giant MC at the systemic velocity of 
$\sim85\km\ps$. This association is strengthened by the
detection of the extended and compact 1720~MHz OH maser emission
and the \HCOp\ emission at $85\km\ps$ from the SNR region.

The southeastern half of the circle is roughly X-ray bright, while
the other half appears to be X-ray faint but covered by the
79--$82\km\ps$ molecular gas. This brightness anticorrelation
between the X-ray and CO emission, however, does not imply that the
molecular gas in the northwestern half obscures the X-rays. We
estimate the hydrogen column density of this gas (in the interval
79--$82\km\ps$) to be $N_{\rm H}\sim1\E{21}\cm^{-3}$. If there were an
X-ray emitting gas, with similar properties to that observed in the
southeast (temperature $kT_x\sim1.6\keV$ and intervening hydrogen
column $2.4\E{22}\cm^{-2}$; Yusef-Zadeh et al.\ 2003), behind this
northwestern molecular gas, then an extra extinction by the
$N_{\rm H}\sim1\E{21}\cm^{-3}$ gas would only cause an $\sim8\%$ decrease
in the 0.5--2~keV X-ray flux and this X-ray emission could have been
observable.
Thus we conclude that the X-ray faintness in the northwest is
not caused by the absorption of the diffuse 79--$82\km\ps$ molecular gas,
whether this gas is connected to the $\sim85\km\ps$ cloud or not.

\subsection{The Dynamics}\label{dynamics}

\begin{figure*}[tbh!]
\centerline{ {\hfil\hfil
\psfig{figure=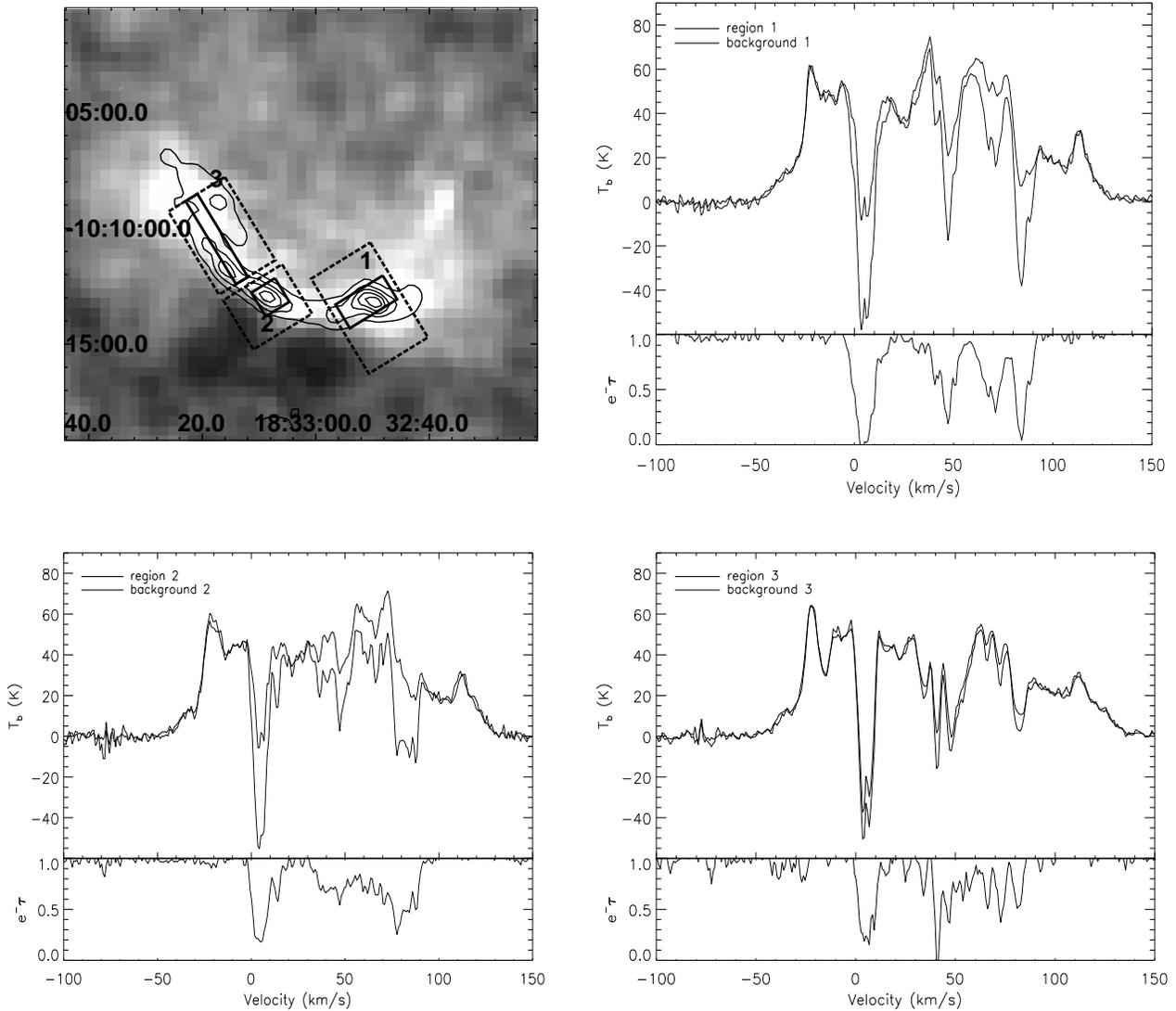,height=6.0in,angle=0, clip=}
\hfil\hfil}} \caption{
{\em Left-top panel}: H{\footnotesize I} emission map of Kes 69 from a single
channel at 74.39 km~s$^{-1}$, overlaid with NVSS radio continuum
contours (with levels at 20, 49, 78, 107, 136 and 165 mJy beam$^{-1}$).
The three pairs of regions are defined to extract the H{\footnotesize I} spectra.
{\em Other three panels}:
the source and background H{\footnotesize I} emission spectra and
the absorption spectra extracted from the defined regions 1, 2,
and 3, respectively.} \label{f:HIspec}
\end{figure*}
The association between SNR~\snr\ and the MC at systemic velocity
$85\km\ps$ facilitates a convincing determination of the kinematic
distance to the remnant. The systemic velocity $85\km\ps$ is
suggestive of two candidate kinematic distances to the SNR/MC
association, 5.2~kpc and 9.6~kpc. The choice can be made with the
aid of the H{\footnotesize I} absorption along the line of sight. Following the
method used in Tian, Leahy, \& Wang (2007) and  Tian \& Leahy
(2008a), we produced three H{\footnotesize I} spectra of regions along the
southeastern shell of \snr, as shown in Figure~\ref{f:HIspec}.
Distinct absorption features appear at 4, 6, 15, 40, 48, 67, 70, 82,
and $88\km\ps$. The \twCO\ components along the shell around 5 and
50 $\km\ps$ (Fig.~\ref{f:regionspec}) can, due to the corresponding
absorption features in the H{\footnotesize I} spectra, can be related to the
chance-coincident foreground gas. No H{\footnotesize I} absorption features are
present around the tangent point velocity $113.7\km\ps$, which
indicates that the SNR is in front of the tangent point (at
7.4~kpc). Hence, the distance to the SNR is $d=5.2\kpc$. We note
that a similar estimate has in the meantime been given by Tian \& Leahy (2008b).

The southeastern molecular arc is revealed above to be coincident
with the radio and IR shell and move at a velocity $v_m$ of order
$10\km\ps$ (matching the blueshifted line broadening which reflects
the velocity component in the line of sight).
The broadened blue
wing of the $\sim85\km\ps$ line profile of the molecular arc implies
that the giant molecular cloud may have suffered a perturbation from
the rear side. Three scenarios regarding the dynamical relation
between the molecular arc and the SNR are discussed below.

First, the arc is likely to be a flake of molecular gas that is
shocked by the slow transmitted cloud shock after the SNR blast wave
hits the molecular cloud. In this case, the radio continuum emission
may arise from the SNR shock that is blocked by dense cloud (Frail
\& Mitchell 1998) or the blast wave that propagates in the
intercloud medium (Blandford \& Cowie 1982), and the thermal X-ray
emission may be ascribed to the hot gas behind the shocked molecular
gas or just behind the blast wave. Thus there can be a crude
pressure balance between the cloud shock and the X-ray emitting hot
gas (Zel'dovich \& Raizer 1967; McKee \& Cowie 1975): $n_0 v_b^2\sim
n_m v_m^2$, where $n_m$ denotes the number density of the hydrogen
atoms ahead of the cloud shock, $n_0$ the density of the undisturbed
intercloud medium, and $v_b$ the velocity of the blast wave.
Velocity $v_b$ is related to the postshock temperature $kT_x$ as
$v_b=[16kT_x/(3\bar{\mu}m_{\rm H})]^{1/2}$, where $m_{\rm H}$ is the hydrogen
atom mass and $\bar{\mu}=0.61$ is the average atomic weight.
Adopting $kT_x\sim1.6\keV$ from the {\sl ROSAT} X-ray observation
(Yusef-Zadeh et al.\ 2003), we have $v_b\sim1.2\E{3}\km\ps$ and
$n_m\sim1.3\E{3}(n_0/0.1\cm^{-3})(v_m/10\km\ps)^{-2}\cm^{-3}$,
implying that the blast wave hits a very dense matter in the
molecular arc (here $n_0$ is assumed to be similar to the mean
density of the X-ray emitting gas, $\sim0.1\cm^{-3}$, as obtained
from a reproduced {\sl ROSAT} X-ray spectral analysis\footnote{ The
mean density of the X-ray-emitting gas is estimated as
$\sim0.14\du^{-1/2}\cm^{-3}$ (where $\du=d/5.2\kpc$ is used for
scaling) from the volume emission measure $n_e n_{rm H}
V\sim7.9\E{57}\du^{2}\cm^{-3}$ of an elliptical region (with
half-axes $11'.2\times6'.7$), for which a volume of an oblate
spheroid (with half-axes $11'.2\times11'.2\times6'.7$) is assumed.
}). In this scenario, assuming an adiabatic expansion and adopting
the SNR extent represented by the circle of radius $\sim8'.7$
(\S\ref{sec:morph}) or $r_s\sim13\du\parsec$, the SNR's age is
estimated as $t=(2r_s)/(5v_b)\sim4.4\du$~kyr and the explosion
energy is
$E=(25/4\xi)(1.4n_0m_{\rm H})v_b^2r_s^3\sim6\E{50}(n_0/0.1\cm^{-3})\du^3
\erg$ (where $\xi=2.026$).

The 4.5$\mu$m and 5.8$\mu$m mid-IR emissions are suggested to be
likely dominated by lines of shocked gas (Reach et al.\ 2006). They
note that the 5.8$\mu$m emission is relatively strong, but it is not
clear whether H$_2$ or [Fe {\footnotesize II}] lines (as likely mechanisms) are
responsible for this emission, unlike the case of IC~443, in which
the [Fe {\footnotesize II}] and H$_2$ lines are very clearly segregated to the
northern and southern regions of the remnant, respectively (see Rho
et al.\ 2001). The H$_2$ emission seems to be consistent with the slow
molecular shock, while [Fe {\footnotesize II}] emission could not be ruled out with
the present observations if the SNR shocks propagate in a
complicated multi-phase medium.
The $24$ $\mu$m emission along the arc is likely to arise from the
shocked molecular gas (e.g., OH and H$_2$O), shock-heated dust
grains, and even probably ions. The OH maser, \HCOp, and H$_2$O (if
any) emissions are consistent with a C-type molecular shock.

Second, let us discuss the scenario in which the arc represents
the interstellar material swept up all the way by the SNR shock
wave. The mass of the southeastern molecular arc is in the range of
0.5--$1.1\E{4}\du^2$ $M_{\odot}$, with the gas mass observed in the
whole line profile adopted as the upper limit and that of the
broadened part as the lower limit (Table~\ref{tb:SE}). In the
swept-up case, the southeastern arc consists of the molecular gas
that was originally of mean density $n({\rm
H}_2)\sim30$--$70\du^{-1}\,{\rm cm}^{-3}$, distributed in an
approximate 1/8th volume of a sphere of radius
$r_s\sim8'.7\sim13\du\parsec$ (considering the one-sided line
broadening). In this scenario, the velocity of the SNR shock is
represented by the arc's slow expansion ($v_m\sim10\km\ps$), which
would imply that the SNR is in the radiative phase. The explosion
energy of the SNR, \(E\sim 7\E{50}(r_s/13\parsec)^{3.12}[n({\rm
H}_2)/50\cm^{-3}]^{1.12}$ $(v_m/10\km\ps)^{1.4}\erg\) (Chevalier
1974), would have seemed to be normal, despite the seldom large age
$t\sim0.31r_s/v_m\sim$ $4\E{5}(r_s/13\parsec)$
$(v_m/10\km\ps)^{-1}\yr$ for X-ray-bright SNRs. However, the X-rays
are bright along the southern shell and two X-ray emission peaks are
essentially coincident with the two radio peaks (Fig.~\ref{f:3c}; 
toward which the pointing molecular line observation was made). A
radiative shock as slow as $10\km\ps$ could not be responsible for
such X-rays along the SNR rim. On the other hand, to accelerate
particles to relativistic energies for emitting radio synchrotron,
the lower limit of the shock velocity is (Draine \& McKee 1993):
$v_s > 64 (x_i/10^{-5})^{1/8}$ $[n({\rm H}_2)/50\cm^{-3}]^{1/8}$
$(\phi_{\rm cr}/0.1)^{-1/4}$ $(T/100\K)^{0.1}\km\ps$, where $x_i$ is
the ionization fraction, $\phi_{\rm cr}$ the efficiency of the
particle acceleration of the shock, and $T$ the preshock molecular
gas temperature. It is much larger than the observed expansion
velocity of the molecular shell, and thus new particle acceleration
would be difficult to take place at this shock. The existing
relativistic electrons would rapidly stream freely away from the
shock front for a very low ionization fraction (Blandford \& Cowie
1982). Moreover, because of the ambipolar diffusion, the magnetic
field will separate from the neutral gas in a timescale (Blandford
\& Cowie 1982): $\sim10^3(\delta r/1\parsec)^2$ $(x_i/10^{-5})$
$[n({\rm H}_2)/50\cm^{-3}]^2$ $(B/10^{-6}\G)^{-2}$ $(s/100)^{-2}\yr$
(where $B$ is the magnetic field strength in the unshocked gas, $s$
the density compression ratio, and $\delta r$ the shell thickness),
very likely to be smaller than the remnant age in this case (even
for $s\sim10$). Therefore, such a slow expansion of molecular cloud
is difficult to account for the SNR's radio emission. In view of the
above points, this scenario does not apply to the molecular shell of
\snr.

Third, the molecular shell may be the debris
of the cooled, condensed material, which was swept up by the stellar wind
of the supernova progenitor from the molecular gas
[$n({\rm H}_2)\sim50\cm^{-3}$] and is now hit by the SNR shock.
This possibility could be compatible with the first scenario
and naturally explain why there is pre-existing, very dense material
that is now hit by the SNR shock.
A molecular wind-bubble shell has recently been discovered coincident
with the ring nebula G79.29+0.46 surrounding a luminous blue
variable star (Rizzo et al.\ 2008), which provides an instance
for this scenario.
If there was a stellar-wind bubble, it was created
(Castor et al.\ 1975; Weaver et al.\ 1977)
$\sim1.4\E{6}(r/13\,{\rm
pc})^{5/3}L_{36}^{-1/3}[n({\rm H}_2)/50\,{\rm cm}^{-3}]^{1/3}$~yr
ago,
where $L_{36}$ is the mechanical luminosity of the stellar wind in
units of $10^{36}\,{\rm erg}\,{\rm s}^{-1}$,
and the present velocity
of the shell is $\sim6(r/13\,{\rm pc})^{-2/3}L_{36}^{1/3}$ $[n({\rm
H}_2)/50\,{\rm cm}^{-3}]^{-1/3}\,{\rm km~s}^{-1}$. This velocity is
comparable to the blueshift in the broadened line profile of the
molecular gas along the shell.

There appears to be a blowout morphology outlined by the
extension of the radio/CO shell to the northeast out of the circle
(Figs.~\ref{f:COregion} and~\ref{f:3c}),
somewhat similar to the blowout morphology in SNR~N132D (e.g.,
Dickel \& Milne 1995; Xiao \& Chen 2008). Actually, SNR~N132D, which
is in the vicinity of an MC, has been suggested to be shaped by the
shock impacting on the stellar wind-bubble shell (Hughes 1987; Chen et
al.\ 2003). The northeastern compact masers at both 69 and
$85\km\ps$ are projected in the blowout region. In this case, as a
possibility, the $69\km\ps$ maser could arise from a dense clump
deviating from the systemic velocity by a strong perturbation,
although it cannot be excluded from being nonassociated with the
same SNR.


\section{Summary}
We have performed a millimeter observation in CO and \HCOp\ lines
toward SNR~\snr.
From the northeastern compact 1720~MHz OH maser region,
the \twCO\ and \thCO\ emission's peaks around $65\km\ps$
and $85\km\ps$, which are consistent with the masers' LSR
velocities are detected.
In the southeast, a molecular (\twCO\ J=1--0) arc is revealed at
77--$86\km\ps$, well coincident with the partial
SNR shell detected in the 1.4~GHz radio continuum and
mid-IR observations. An $85\km\ps$ \HCOp\ emission is found to arise
from a radio peak on the shell. Both the molecular arc and the
\HCOp\ emission at $\sim85\km\ps$ seem to be consistent with the
presence of the extended 1720 MHz OH emission along the southeastern
boundary of \snr. The morphology correspondence between the CO
emission and other band emission of the \snr\ shell provides strong
evidence for the association of the SNR with the $\sim85\km\ps$
component of molecular gas. There is another section of molecular
arc at 79--$86\km\ps$ in the northwest.  Both the molecular arcs,
together with the faint northern radio features, seem to be distributed
along a circle of radius $8'.7$.
The multiwavelength emissions along the southeastern shell can be
explained by the impact of the SNR shock on a dense, clumpy patch of
molecular gas. This pre-existing gas is likely to be a part of the
cooled, clumpy debris of the interstellar molecular gas swept up by the
progenitor's stellar wind.  The association
between SNR~\snr\ and the MC at the systemic velocity $\sim85\km\ps$
enables us to place the SNR at a kinematic distance of 5.2~kpc.
\\

{\sl Note added in proof.} In another observation in CO lines toward SNR Kes 75,
we similarly discover a molecular shell, a part of which follows the bright
partial SNR shell seen in X-rays, mid-IR, and radio continuum, and provide effective
evidence for the association between Kes 75 and the adjacent MC (Su et al. 2009).
Molecular shells are probably common in a number of SNRs but have rarely been studied.

\acknowledgments We are grateful to the staff of Qinghai Radio
Observing Station at Delingha for the support during the
observation. We thank Wen-Wu Tian for advice on the distance
estimates and other specific issues. Y.C.\ acknowledges support
from NSFC grants 10725312, 10673003, and 10221001 and the 973
Program grant 2009CB824800. We acknowledge the use of the VGPS data;
the National Radio Astronomy Observatory is a facility of the
National Science Foundation operated under cooperative agreement
by Associated Universities, Inc.

\end{document}